\documentclass[prl,aps,superscriptaddress,twocolumn,floatfix,amssymb]
{revtex4}
\usepackage{epsfig}

\newcommand{\be}{\begin{equation}}
\newcommand{\ee}{\end{equation}}
\newcommand{\bc}{\begin{center}}
\newcommand{\ec}{\end{center}}
\newcommand{\bi}{\begin{itemize}}
\newcommand{\ei}{\end{itemize}}
\newcommand{\ba}{\begin{eqnarray}}
\newcommand{\ea}{\end{eqnarray}}

\newcommand{\ignore}[1]{}
\newcommand{\ie}{{\em i.e.\ }}
\renewcommand{\d}{{\rm d}}

\begin{document}

\title{Scaling in the structure of directory trees in a computer cluster}

\author{Konstantin Klemm}
\affiliation{Interdisciplinary Centre for Bioinformatics /
Bioinformatics Group of the Institute for Computer Science,
University Leipzig, Kreuzstr.~7b, 04103 Leipzig, Germany}
\author{V\'{\i}ctor M. Egu\'{\i}luz}
\affiliation{Instituto Mediterr\'aneo de Estudios Avanzados IMEDEA
(CSIC-UIB), E07122 Palma de Mallorca (Spain)}
\author{Maxi San Miguel}
\affiliation{Instituto Mediterr\'aneo de Estudios Avanzados IMEDEA
(CSIC-UIB), E07122 Palma de Mallorca (Spain)}
\date{\today}

\begin{abstract}

We describe the topological structure and the underlying organization
principles of the directories created by users of a computer cluster when
storing his/her own files. We analyze degree distributions, average
distance between files, distribution of {\em communities} and {\em
allometric scaling} exponents of the directory trees. We find that users
create trees with a broad, scale-free degree distribution. The structure
of the directories is well captured by a growth model with a single
parameter. The degree distribution of the different trees has a
non-universal exponent associated with different values of the parameter
of the model. However, the distribution of community sizes has a
universal exponent analytically obtained from our model.
\end{abstract}
\maketitle

The processes of storing and retrieving information are rapidly gaining
importance in science as well as society as a whole
\cite{Shiffrin04,Lawrence99,Ferrer01,Sigman02}. A considerable effort is being undertaken,
firstly to characterize and describe how publicly available information,
for example in the world wide web, is actually organized, and secondly,
to design efficient methods to access this information. It seems clear
that to design methods for accessing information we first need to know how
information is actually stored or organized as it is being produced.

Within this general framework a crucial step in building general
knowledge on these processes, is the understanding of how each of us
organizes knowledge and information produced by ourselves. To be
specific, we pose the question of general organizational principles in
the managing of our own electronic files. To answer this question we
analyze the structure and organization of the files stored in a computer
cluster by the users of the computer facilities at a research institute.
Within the general study of complex networks, we are here looking at {\em
trees} and we report a first observation of the scale free property in
trees. It is important to point out that we are not studying a
single large tree, but rather we are considering a forest of many trees,
each of them being the result of an individual construction. We are then
able to consider samples of organizational schemes of many different
sizes, since each user has created a structure with a different number of
directories. This allows the study of different samples of the same reality.
We also note that contrary to
other networks like the WWW or food webs, the structures considered here
are not the outcome of a collective action but the creation of a single
individual. Our research gives information about the management of
information at the individual level.

Two {\it a priori} possible answers to the question posed are that we
follow a {\em random} process of file storing or that, on the contrary,
we implement a careful {\em planned} structure as we do when organizing
the sections and chapters of a PhD thesis or a scientific paper. What we
find is the signature of a complex system halfway between these two
possibilities, but still with well defined patterns of organization. In
this paper we report an extensive characterization of individual user
computer directory trees, calculating a number of quantitative measures.
These include degree distributions, average distance between directories
\cite{Strogatz01,Albert02,Dorogovtsev02,Newman03r}, distribution of
community sizes in the tree \cite{Guimera03} and allometric scaling
exponents \cite{Banavar99,Garlaschelli03}. Our data turns out to be well
described by a directory attachment model for constructing the tree. The
model depends on a single parameter $q$ that interpolates between random
placement of new directories and the agglomeration into a star structure.
The trees of the different users are described by different values
of the parameter $q$: diversity in individual behavior here boils down to
a different value of a parameter.

\begin{figure}
\centerline{\epsfig{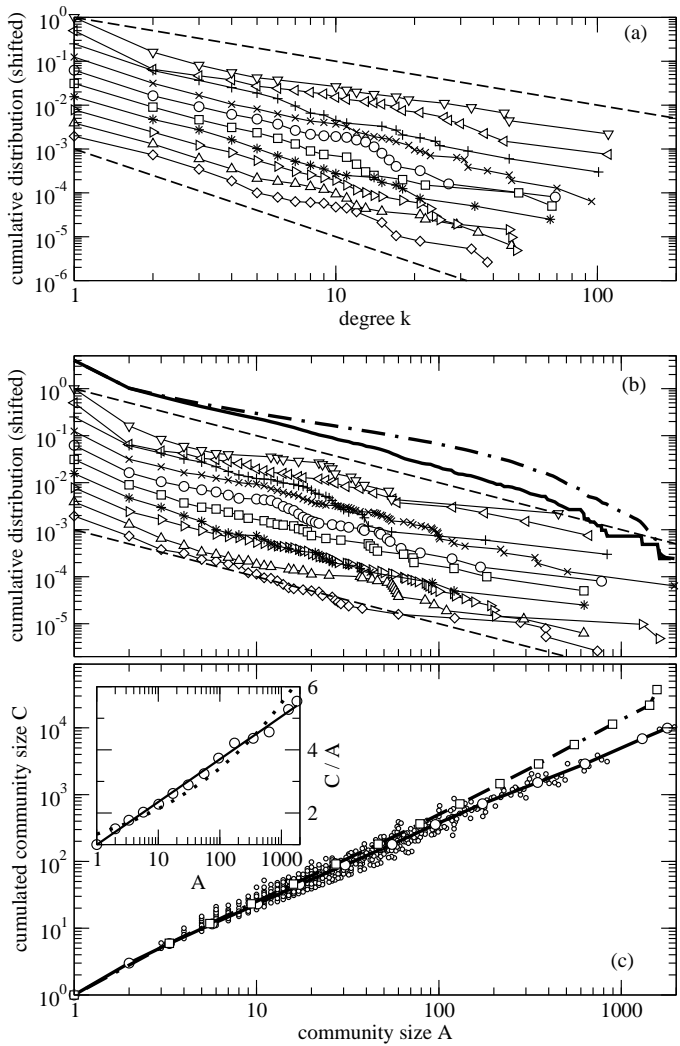}}
\caption{\label{fig1} Scaling in the distributions of branching
ratio (degree) and sizes of the communities (subtrees). (a)
Cumulative degree distributions for the ten largest trees. The
dashed lines have slopes -1 and -2 indicating degree exponents
between $\gamma=2$ and $\gamma=3$. In the whole data set, however,
exponents $\gamma>3$ have been observed as well. (b) Cumulative
distributions of community size plotted as in (a). The dashed
lines have slope -1 corresponding to community size exponent $\tau
= 2$. The overall cumulative distribution of the sizes of the
16452 communities in all 63 directory trees (thick solid curve) and the
surrogate data from randomized trees
(dot-dashed curve) are shown as well. (c) Allometric scaling: Each
data point (small circle) shows cumulative community size $C$ (sum of
sizes of all sub-communities) against the size $A$ of the community
itself. Logarithmic binning is applied to the original data
(large circles) and the surrogate data from randomized trees
(squares). The inset shows the binned original data rescaled with
$A$ (circles) and best fits for logarithm (solid line) and power law
(dotted curve). The surrogate data in (b) and (c) are taken from
6300 trees, 100 trees obtained from
each original tree by independent random rewiring. Rewiring is
performed by iteratively swapping two randomly chosen node
disjoint subtrees that do not contain the root. This standard
network randomization procedure \cite{Maslov02}, here applied to
rooted trees, conserves the degree distribution. }
\end{figure}

{\em Data analysis.}-- The data material under consideration is taken
from the computer facilities of the Cross-disciplinary Physics Department
of IMEDEA (Mediterranean Institute for Advanced Studies). The personal
accounts of the 63 users running Linux and UNIX have been considered. The
users include academic staff, post-docs, graduate students and long-time
visitors. Each user is able to choose freely his/her own organizational
scheme without specific software. The nodes in the directory tree of a
given user are all directories (file folders) stored in the user's
computer account. There is a direct link between nodes $i$ and $j$ if
directory $i$ is a subdirectory of directory $j$ or vice versa. We
consider the trees as rooted with the home directory as the root. In the
following, we analyze the trees in terms of the distributions of degree
and of community sizes as well as the allometric scaling.

A local measure of the importance of a given node $i$ is the nodal degree
$k_i$ counting the number of nodes directly connected to $i$. In a tree
of $N$ nodes the average degree is always $\langle k \rangle = 2 - 2/N$.
The distribution of the degree, however, varies strongly across different
types of structures. The distribution is narrow in simple chains and
binary trees while it is broadest for a star (having $N-1$
nodes with degree $k=1$ and one center node with degree $k=N-1$). The
degree distributions of the observed directory trees (Fig.\
\ref{fig1}(a)) lie in between these two extremes. Directory trees are
scale-free. The probability of finding a node with degree $k$ decays as a
power law $k^{-\gamma}$ with a cut-off at the maximum degree $k_{max}$
due to finite size. There is no indication of an upper bound on the
degree that would limit the scaling at large $k$. Given trees generated
by different users, the observed values of $\gamma$ do not coincide in
general. The degree exponent is not universal.

An alternative characterization of the trees is obtained by iterative
decomposition into subtrees rather than single nodes. Here we consider
the {\em community structure} of the trees. For each node $i$, a
community $S_i$ is the subtree rooted at the node $i$ and all nodes
below $i$. In the directory trees, a community $S_i$ is the tree formed
by a directory $i$, all its subdirectories, the subdirectories of these
and so forth. A community $S_i$ is again a rooted tree with node $i$ as
the root. Calculating the sizes $A_i=|S_i|$ of all communities for each tree,
we find the statistics in Fig.\ \ref{fig1}(b). The distribution of
community sizes decays as a power law $A^{-\tau}$. The exponent
$\tau = 2$ appears to be universal. The scaling of community size
$A$ is a property independent of the scaling of the degree $k$. When the
trees are randomized under conserving degrees of all nodes, the
functional form of the community size distribution changes and obtains a
scaling region with a larger exponent $\tau>2$.

In order to capture also the correlations between community sizes we
perform {\em allometric scaling} analysis \cite{Garlaschelli03}.
For each community $S_i$ we 
calculate the quantity $C_i = \sum_{j\in S_i} A_j$, \ie we sum up all
the sizes of all communities contained in $S_i$, including $S_i$ itself.
Figure \ref{fig1}(c) shows the data point $(A_i,C_i)$ for each community
$i$ in the 63 trees. We find that the growth of $C$ with $A$ is
superlinear.

\begin{figure}
\centerline{\epsfig{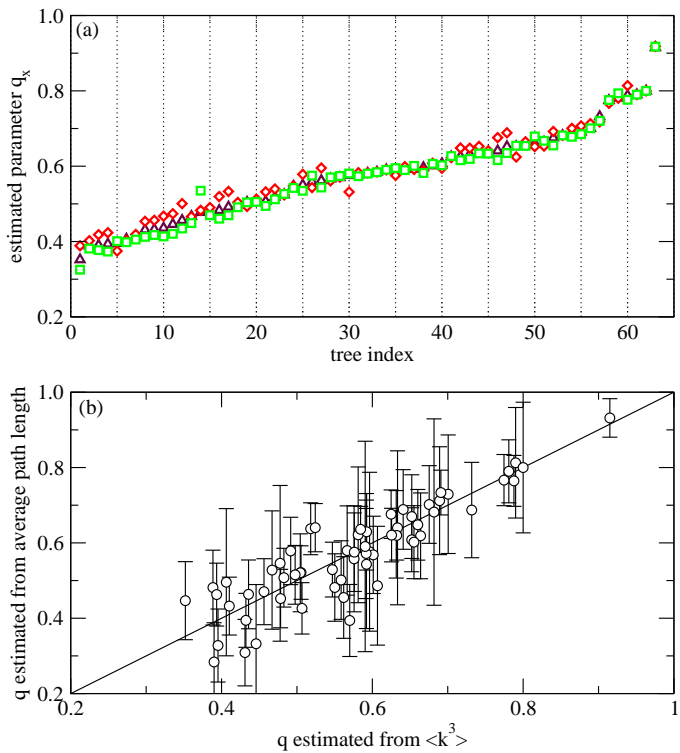}}
\caption{\label{fig:modelf}
Estimating the model parameter $q$ from the empirical trees.
(a) Independent estimates of $q$ from the moments $\langle k^n \rangle$ of
the degree distribution coincide for each tree. Estimates are plotted for
$n=2$ (diamonds), $n=3$ (triangles), and $n=4$ (squares). Tree index
reflects the ordering of the trees with respect to the estimated $q$.
(b) Comparing the $q$ values estimated from the average path length and
the third moment of the degree for each tree. For all estimates in (a)
and (b) the following method is used. Given an empirical tree of size $N$
with observable $x_{\rm emp}$, $10^5$ parameter
values $q\in[0,1]$ are drawn equally distributed. For each value drawn an
artificial tree of size $N$ is generated by the model. The tree is accepted
if its value $x_{\rm model}$ of the considered observable does not differ
by more than $10\%$ from the empirical value $x_{\rm emp}$. We take
$\langle q \rangle$ as the average over parameter values of all accepted
trees. The range of an error bar in (b) indicates the standard deviation
of $q$ across the accepted trees.
}
\end{figure}

{\em Modeling.}-- Let us now consider a stochastic model for the
construction of a directory tree. We assume that users build their trees
by iteratively adding nodes, \ie creating new directories. Then for each
possible tree the model assigns an attachment probability to each of the
nodes. The attachment probability $\Pi_i$ of a node $i$ is the
probability that $i$ becomes the parent of the next added node. In the
simplest case, the structure of the tree is irrelevant for the attachment
process. Then we have {\em homogeneous attachment}. Each directory has
the same probability to become the parent of a new directory. Another
conceivable rule is {\em copying} of directories. If directories are
chosen for duplication with equal probability, a directory obtains a new
subdirectory with a probability proportional to the number of
subdirectories it already has. Here we formulate a model comprising both
these mechanisms at tunable ratio. In a tree with $N$ nodes, a node with
degree $k$ becomes the parent of the next added node with probability
\begin{equation} \label{eq:attachrate}
\Pi(k) = q \frac{k-1}{N} + (1-q) \frac{1}{N}~.
\end{equation}
The tunable parameter $q\in[0,1]$ is the probability that duplication of
a node is performed. With probability $1-q$ a randomly chosen node is the
parent of the added directory. Qualitatively $q$ measures how often the
individual creating the tree likes to subdivide a directory. Note that
the pure rules ($q=0$, $q=1$) cannot produce trees as in Fig.\
\ref{fig1}(a). Homogeneous attachment ($q=0$) leads to trees with an
exponential degree distribution. The pure duplication mechanism ($q=1$)
can only generate stars because it cannot turn a leaf into an inner node.
By rewriting Eq.\ (\ref{eq:attachrate}) as $\Pi(k) \propto k^{\rm in} +
a$ with the number of links $k^{\rm in} = k -1$ received after creation
of the node and the ``initial attractiveness'' $a = 1/q -1$ we see the
equivalence of our model with the network growth model by Dorogovtsev et
al.\ \cite{Dorogovtsev00}, restricted to a single new link added per
node. The case $q=1/2$, giving $\gamma=3$, is the
scale-free model by Barabasi and Albert \cite{Barabasi99}.
For general $q\in[0,1[$, the model produces scale-free trees with
degree exponent $\gamma = 2+a = 1+1/q >2$.

The evolution of community sizes is described by the probability
\begin{equation} \label{eq:commrate}
{ \tilde \Pi} (A) = q \frac{A-1}{N} + (1-q) \frac{A}{N} = \frac{A-q}{N}~.
\end{equation}
that the next node is attached to one of the nodes of a given community
of size $A$, thereby incrementing $A$. From a continuous rate equation approach
\cite{Barabasi99} we obtain $A_i (N) = (1-q) N/i + q$ as the expected
size of community $S_i$ in a tree of size $N$. The index $i$ is the time
step of creation of the community as a single node with $A=1$. 
The linear growth of $A$
with $N$ implies that the community size distribution of the model decays
asymptotically as $A^{-\tau}$ with universal ($q$-independent) exponent
$\tau=2$, in agreement with the data.

For an estimate of the allometric scaling, first note the general
property $C_i = A_i + \sum_{j\in {S_i}} d_{ij}$ where the chemical
distance $d_{ij}$ is the number of nodes contained in the direct path
between nodes $i$ and $j$. Adding a new node $j^\ast$ to community $S_i$,
the expected distance $\langle d_{ij^\ast} \rangle$ from node $i$ is $C_i
/ A_i - 1$ for copying and $C_i / A_i$ for homogeneous attachment. Thus
on average $C$ grows as $\d C / \d A = 1 + C / A - q$, where the finite
difference has been approximated by the derivative and the index $i$ is
suppressed. For the initial condition $C (1) = 1$ we obtain the solution
$C(A) = A[(1-q)\ln A + 1]$. The allometric scaling of the model trees is
linear with logarithmic correction. In order to compare with the observed
trees we replot the binned data as $(A_i,C_i/A_i)$ in the inset of Fig.\
\ref{fig1}(c). The data are captured well by a logarithmic dependence
(best fit $C/A = 0.59 \ln A + 0.99$, correlation coefficient $r=0.997$)
in good agreement with the prediction of the model.


In order to provide a more stringent check of the validity of the model
(Eq.\ (\ref{eq:attachrate})) we first project the trees into a space of
four observables, namely the second, third and fourth moments of the
degree distribution and the average chemical distance between nodes. For
a given value $x$ of an observable and given tree size $N$ we estimate
the most likely parameter value $q_x$ by weighting all possible values
$q\in[0,1]$ with the probability that they produce $x$ up to a small
error. Figure \ref{fig:modelf} shows the results and gives details of the
method in the caption. For almost all trees there is excellent agreement
between the four parameter estimates based on different observables. Thus
after choice of a single parameter the model accurately reproduces the
projection of the trees into a four-dimensional space. The projection
takes into account the distribution of the degree as a local property,
and the average distance $\langle d_{ij} \rangle$ between nodes as a
global property. This is strong evidence that the proposed growth
mechanism produces statistically the same structures as seen in the
directory trees.

{\em Discussion.}-- The structure of directory trees has been
characterized from a statistical point of view. Our main result is the
striking structural similarity between trees created by independent users
in the absence of common constraints. Users create trees with a broad,
scale-free degree distribution with a non-universal exponent. The
distribution of community sizes, however, scales with a universal
exponent $\tau \approx 2$. The allometric scaling is linear with a
logarithmic correction. Community structure and allometric scaling are
significantly different in random surrogate trees with the same degree
distribution. The statistical properties of the empirical trees are
reproduced by a model that generates trees by adding nodes iteratively.
The model has a single parameter $q$ controlling the tendency to
accumulate many subdirectories in the same parent directory.
By varying $q$, the degree exponent can be tuned in the
empirically observed range $\gamma$. The exponent $\tau =2$ and the
allometric scaling $ C \sim A \ln A$ have been derived analytically and
are independent of the parameter $q$. The validity of the model has been
evidenced further by determining the most likely value of the parameter
$q$. For a given tree, estimates based on different moments of the degree
distribution as well as the diameter coincide, while estimates vary
across trees. Consequently, directory trees can be distinguished by their
specific value of the growth parameter $q$.

A generally interesting question is to decide about universal properties
and universality classes of different natural and artificial or man-made
complex networks. The community distribution exponent $\tau \simeq 2$
that we find for our directory trees is in agreement with the one
reported for the Internet \cite{Caldarelli00,Rios01} and for the
communities of scientific collaborations \cite{Radicchi04,Arenas03a}. 
However a different class is formed by river networks
\cite{RodriguezIturbe96,Rinaldo93,Maritan96}, informal networks in
organizations \cite{Guimera03} and jazz musician networks
\cite{Arenas03a}, where the corresponding exponent gives a value $\tau
\sim 1.45$ \cite{Rios01}. These examples seem to belong to the class of
efficient networks obtained from an optimization principle in which
transportation costs are minimized \cite{Banavar99}. For the class of
efficient networks one can prove \cite{Banavar99,West97,Banavar02} that
allometric scaling is given by a a power law dependence $C \sim
A^{\eta}$, with a universal exponent $\eta = (D+1)/D$, where $D$ is the
embedding dimension. At difference with the prediction from efficiency,
we find $C \sim A \ln A$ for the directory trees as reproduced by our
growth model. This result is also compatible with effective (apparent)
exponents observed in food webs \cite{Garlaschelli03}.

We have shown that directory trees as individually man-made but not designed
objects are an interesting direction of further research into hierarchical
networks. Analyzing the wealth of readily available tree data on computers
around the world offers improved insight into how people naturally structure
information.

We acknowledge financial support from MEC (Spain) through project
CONOCE2 (FIS2004-00953) and from Deutsche Forschungsgemeinschaft (DFG).


\end{document}